\documentclass[superscriptaddress,groupedaddress,nofootnoteinbib,12pt]{article} 
\pdfoutput=1
\usepackage{graphicx}
\usepackage{color}
\usepackage{dcolumn}   
\usepackage{bm}       
\usepackage{amssymb}  
\usepackage{amsmath}
\usepackage{sectsty}
\usepackage{latexsym}
\usepackage{float}
\usepackage{ifthen}
\usepackage{caption,subfig}
\usepackage{enumerate}
\usepackage{url}
\usepackage{caption,subfig}
\usepackage{feynmf}	
\usepackage{jcappub}
\usepackage{amsopn}
\usepackage{float}
\usepackage{pst-pdf}
\usepackage{amsfonts}
\usepackage{multirow}
\usepackage{array}
\usepackage{booktabs}
\usepackage{rotating}
\usepackage{color}

\def\clap#1{\hbox to 0pt{\hss#1\hss}}

\def\({\left(}
\def\){\right)}
\def\[{\left[}
\def\]{\right]}
\def\bea{\begin{eqnarray}}
\def\eea{\end{eqnarray}}
\def\be{\begin{equation}}
\def\ee{\end{equation}}
\def\ba{\begin{eqnarray}}
\def\ea{\end{eqnarray}}
\def\beq{\begin{eqnarray}}
\def\eeq{\end{eqnarray}}
\def\mpl{M_{\rm Pl}}

\def\clap#1{\hbox to 0pt{\hss#1\hss}}

\definecolor{forestgreen}{rgb}{0.133,0.545,0.133}
\newboolean{editorial}
\setboolean{editorial}{true}
\newcommand{\editorial}[2]{\ifthenelse{\boolean{editorial}}{\textcolor{red}{[\textsf{\textbf{{#1}}}: }\textcolor{blue}{\textsf{{#2}}}\textcolor{red}{]}}{}}

 \def\be   {\begin{equation}}   \def\ee   {\end{equation}}

 \def\ba  {\begin{eqnarray}}   \def\ea  {\end{eqnarray}}

\renewcommand{\thesubsection}{\arabic{section}.\arabic{subsection}}

\begin{document}

\title{Cosmology of the proxy theory to massive gravity}

\author{Lavinia Heisenberg,$^{a,b}$,
Rampei Kimura $^{c}$ and Kazuhiro Yamamoto$^{d}$}
\affiliation{$^{a}$Perimeter Institute for Theoretical Physics, \\
31 Caroline St. N, Waterloo, Ontario, Canada, N2L 2Y5}
\affiliation{$^{b}$D\'epartment de Physique  Th\'eorique and Center for Astroparticle Physics,\\
Universit\'e de Gen\`eve, 24 Quai E. Ansermet, CH-1211  Gen\`eve, Switzerland}
\affiliation{$^{c}$Research Center for the Early Universe, \\
The University of Tokyo, Tokyo, 113-0033, Japan}
\affiliation{$^{d}$
Department of Physical Science, Hiroshima University, \\
Higashi-Hiroshima,
Kagamiyama 1-3-1, 739-8526, Japan
}
	
	\emailAdd{Lavinia.Heisenberg@unige.ch}
	\emailAdd{rampei@resceu.s.u-tokyo.ac.jp}
	\emailAdd{kazuhiro@hiroshima-u.ac.jp}

\abstract{
In this paper, we scrutinize very closely the cosmology in the proxy theory to massive gravity obtained in Phys. Rev. D84 (2011) 043503.  
This proxy theory was constructed by covariantizing the decoupling limit Lagrangian of massive gravity and represents a subclass of Horndeski scalar-tensor theory. Thus, this covariantization unifies two important classes of modified gravity theories, namely massive gravity and Horndeski theories.
We go beyond the regime which was studied in Phys. Rev. D84 (2011) 043503 and show that the theory does not admit any homogeneous and isotropic self-accelerated solutions.
We illustrate that the only attractor solution is flat Minkowski solution, hence this theory is less appealing as a dark energy model. We also show that the absence of de Sitter solutions
is tightly related to the presence of shift symmetry breaking interactions.
}


\maketitle


\section{Introduction}

Whether the law of gravitation at cosmological distances can 
be described by general relativity or not will provide us 
a rich information of dark energy,
which is responsible for the present accelerated expansion of the universe.
One of such a candidate for alternative theories of gravity is massive
gravity, originally proposed by Fierz and Pauli \cite{FP:1939aa}. 
They introduced a mass term in the linearized theory of general relativity
in the context of Lorentz invariant theory.
Unfortunately once Fierz-Pauli massive gravity is extended to a nonlinear theory
the sixth degree of freedom called Boulware-Deser ghost appears \cite{Boulware:1972aa}.
This problem was recently solved by de Rham and Gabadadze 
by adding higher order potentials,
making the sixth degree of freedom removed \cite{deRham:2010ik}.
It turned out that this infinite potential can be resummed 
by introducing the tensor,
which has a square-root structure \cite{Rham:2011aa},
and this theory is now referred as de Rham-Gabadadze-Tolley (dRGT) massive gravity, which has been shown to be technically natural \cite{deRham:2012ew, deRham:2013qqa}.
Since the inception of the dRGT theory there has been 
a flurry of investigations related to the self-accelerating solutions 
in the full theory. 
In dRGT theory, the universe can not be of a flat or closed Friedmann-Robertson-Walker (FRW) form \cite{PhysRevD.84.124046},
nonetheless, open FRW universe is still allowed \cite{Gumrukcuoglu:2011aa}.
In this solution, the mass term exactly behaves as the cosmological constant,
which allows a self-accelerating universe. 
However, the perturbations suffer from the instabilities, 
and the kinetic terms in the scalar and vector sector
vanishes, which signals the strong coupling at a certain scale 
\cite{Gumrukcuoglu:2012aa,Gumrukcuoglu:2012ab,Felice:2012aa}.
On the other hand, in Ref. \cite{Rham:2011ab} it has been shown, 
that there are exact de Sitter solutions in the decoupling limit theory,
which is only valid within a certain region in the universe. 
This solution however suffers from ghost instabilities of the vector modes unfortunately
\cite{Rham:2011ab, Koyama:2011wx, Gabadadze:2013aa}.
In any case, it is very interesting that the mass of the graviton
can drive an accelerated expansion of the universe \cite{Leon:2013qh}.

As an alternative of massive gravity, 
one can covariantize the decoupling limit theory \cite{deRham:2011by},
and this "proxy theory" is not a massive gravity theory any longer but rather  
a non-minimally coupled subclass of Horndeski scalar-tensor theory \cite{Horndeski:1974aa}.
Horndeski theory is the scalar-tensor theory, whose equations of motion
remain second order differential equations,
while the Lagrangian contains second derivatives with respect to
space-time. 
It has been shown that Horndeski theory is equivalent with 
generalized Galileon theory \cite{Deffayet:2011aa}, 
which is the general extension of the Galileon theory \cite{Nicolis:2009aa},
and these theory contains four arbitrary functions in the Lagrangian\footnote{See Ref. \cite{Tasinato:2014eka, Heisenberg:2014rta} for the generalized vector Galileons.}. In the proxy theory these arbitrary functions can be automatically 
determined by covariantization, 
and it shares the same decoupling limit with dRGT massive gravity. 
In Ref. \cite{deRham:2011by}, 
the authors found a self-accelerating solution in a given approximated regime
driven by the scalar field ,
which originally represents the helicity-0 mode in massive gravity.
In contrast to the pure Galileon models, generalized Galileons do not impose the Galileon symmetry. 
The naive covariantization of the Galileon interactions on non-flat backgrounds breaks the Galileon symmetry explicitly, however one can successfully generalize the Galileon interactions to maximally symmetric backgrounds while keeping the corresponding symmetries \cite{Burrage:2011bt}. Inspired by these Horndeski scalar-tensor interactions, one can in a similar way construct the most general vector-tensor interactions with non-minimal couplings with only second order equations of motion \cite{Tasinato:2014eka, Heisenberg:2014rta}. The cosmology of these theories has been explored in \cite{Jimenez:2013qsa}.

In the present paper,
we study the cosmological evolution in the proxy theory in more detail beyond the approximations used in \cite{deRham:2011by}
and show the absence of de Sitter attractor solutions 
which renders the theory not suitable as dark energy model.
In Sec.2, we briefly review dRGT massive gravity 
and the derivation of the proxy theory.
In Sec.3, we first investigate the de Sitter solution, then
we study the dynamical system of cosmological solutions 
by using phase analysis. 
In Sec 4, we summarize our results.

Throughout the paper, we use units in which the speed
of light and the Planck constant are unity, 
$c=\hbar=1$, 
and $M_{\rm Pl}$ is the reduced Planck mass related 
to Newton's constant by $M_{\rm Pl}=1/\sqrt{8 \pi G}$. 
We follow the metric signature convention $(-,+,+,+)$.
Some contractions of rank-2 tensors are denoted by
${\cal K}^{\mu}_{~\mu}=[{\cal K}]$,~
${\cal K}^{\mu}_{~\nu}{\cal K}^{\nu}_{~\mu}=[{\cal K}^2]$,~
${\cal K}^{\mu}_{~\alpha}{\cal K}^{\alpha}_{~\beta}{\cal K}^{\beta}_{~\mu}=[{\cal K}^3]$, 
and so on.

\section{Proxy theory to massive gravity}

\subsection{dRGT massive gravity and the decoupling limit}
In massive gravity, one has to introduce the fluctuation tensor $h_{\mu\nu}$,
which measures the mass of graviton, and it is usually defined by 
the difference between the physical metric and the 
Minkowski metric, 
$h_{\mu\nu}=g_{\mu\nu}-\eta_{\mu\nu}$.
Once we introduce a mass term in a gravitational theory, 
the theory does not preserve the diffeomorphism invariance;
however, the diffeomorphism invariance can be restored by introducing 
the St{\" u}ckelberg field $\phi^a$ \cite{Arkani-Hamed:2003aa},
through the relation, 
$H_{\mu\nu}=g_{\mu\nu}-\eta_{ab}\partial_{\mu}\phi^a\partial_{\nu}\phi^b$,
where $H_{\mu\nu}$ is the covariant version of the fluctuation tensor $h_{\mu\nu}$\footnote{The choice of the St{\" u}ckelberg field is arbitrary, and 
fixing the unitary gauge, $\phi^a=\delta^a_\mu x^\mu$, 
$H_{\mu\nu}$ reduces to the orignal fluctuation tensor $h_{\mu\nu}$.}.
Then the action for massive gravity is in general given by
\begin{eqnarray}
  S_{\rm MG}={M_{\rm Pl}^2\over 2}\int d^4x \sqrt{-g}\biggl(R
    -{m^2 \over 4} {\cal U}(g,H)
  \biggr)+S_m(g_{\mu\nu}, \psi),
\end{eqnarray}
where $m$ is the mass of graviton, $U(g, H)$ is the potential terms,
and $S_m$ is the action for the matter fields $\psi$ living on the geometry.
The candidate of potential terms is the Fierz-Pauli mass term,
which is the ghost-free term at quadratic order in $H_{\mu\nu}$  \cite{FP:1939aa}.
However, this term produces a extra ghostly degrees of freedom at
nonlinear level found by Boulware and Deser \cite{Boulware:1972aa}.
In order to eliminate this Boulware-Deser ghost, 
one has to add the infinite nonlinear corrections
in addition to the quadratic potential \cite{deRham:2010ik}.
These infinite nonlinear potentials can be remarkably simplified by using 
the new tensor 
${\cal K}^{\mu}_{~\nu} =\delta^{\mu}_{~\nu} -\sqrt{\delta^{\mu}_{~\nu} -H^{\mu}_{~\nu} }=\delta^{\mu}_{~\nu} -\sqrt{\eta_{ab}g^{\mu\alpha}\partial_{\alpha}\phi^a\partial_{\nu}\phi^b}$, 
and then the resummed potential for ghost-free massive gravity 
is given by \cite{Rham:2011aa}
\begin{eqnarray}
 {\cal U}(g,H)
=-4\left( {\cal U}_2+ \alpha_3 {\cal U}_3 + \alpha_4 {\cal U}_4\right),
\end{eqnarray}
where $\alpha_{3,4}$ are model parameters and 
\begin{eqnarray}
  {\cal U}_2&=&
  -{1\over 2}\varepsilon_{\mu\alpha\rho\sigma}\varepsilon^{\nu\beta\rho\sigma}
  {\cal K}^{\mu}_{~\nu}{\cal K}^{\alpha}_{~\beta}
  =  [{\cal K}]^2-[{\cal K}^2],
  \nonumber\\
  {\cal U}_3&=&
  -\varepsilon_{\mu\alpha\gamma\rho}\varepsilon^{\nu\beta\delta\rho}
  {\cal K}^{\mu}_{~\nu}{\cal K}^{\alpha}_{~\beta}{\cal K}^{\gamma}_{~\delta}
  =  [{\cal K}]^3-3[{\cal K}][{\cal K}^2]
  +2[{\cal K}^3],
  \nonumber\\
  {\cal U}_4&=&
  -\varepsilon_{\mu\alpha\gamma\rho}\varepsilon^{\nu\beta\delta\sigma}
  {\cal K}^{\mu}_{~\nu}{\cal K}^{\alpha}_{~\beta}
  {\cal K}^{\gamma}_{~\delta}{\cal K}^{\rho}_{~\sigma}
  =[{\cal K}]^4-6[{\cal K}]^2[{\cal K}^2]
  +3[{\cal K}^2]^2+8[{\cal K}][{\cal K}^3]-6[{\cal K}^4].
\end{eqnarray}
The sixth degree of freedom is absent in massive gravity with this potential, 
and this theory has five degrees of freedom, which are 
the proper degrees of freedom in massive gravity \cite{Hassan:2011hr, deRham:2011rn}.
The five polarization modes in the ghost-free massive gravity
can be decomposed into the scalar, vector, and tensor modes
by taking the decoupling limit,
which is very convenient to capture the dynamics of each modes 
within the scale $m^{-1}$.
In order to decompose these modes, 
we usually expand the St{\" u}ckelberg field around
the unitary gauge\footnote{The vector modes are disregarded for simplicity. 
For the detail of complete derivation, see \cite{Ondo:2013wka, Gabadadze:2013aa}.} as
\begin{eqnarray}
  \phi^a=\delta^a_\mu x^\mu -\eta^{a\mu} {\partial_\mu \pi / M_{\rm Pl}m^2},
\label{stuckelbergDL}
\end{eqnarray}
and the physical metric around Minkowski background as
$g_{\mu\nu}=\eta_{\mu\nu}+h_{\mu\nu}/M_{\rm Pl}$,
where $\pi$ describes the scalar mode of a massive graviton. 
Then the decoupling limit can be taken by the following limits,
\begin{eqnarray}
  M_{\rm Pl} \to \infty, \qquad m \to 0, 
  \qquad \Lambda_3 = (M_{\rm Pl}m^2)^{1/3}={\rm fixed}.
\end{eqnarray}
The Lagrangian in the decoupling limit takes the following simple form
\begin{eqnarray}
\mathcal{L}=-\frac{1}{2}
h^{\mu\nu}\mathcal{E}^{\alpha\beta}_{\mu\nu} h_{\alpha\beta}+
 h^{\mu\nu}\sum_{n=1}^3 \frac{a_{n}}{\Lambda_3^{3(n-1)}} X^{(n)}_{\mu\nu}[\Pi]
+\frac{1}{2\mpl}h^{\mu\nu}T_{\mu\nu},
\label{lagr}
\end{eqnarray}
where the first term represents the usual kinetic term for the
graviton defined in the standard way with the Lichnerowicz operator given by 
\begin{equation}
{\mathcal{E}}^{\alpha\beta}_{\mu\nu} h_{\alpha\beta}=-\frac 12 \left(\Box h_{\mu\nu}-2\partial_\alpha \partial_{(\mu}h^\alpha_{\nu)}+\partial_\mu\partial_\nu h-\eta_{\mu\nu} (\Box h-\partial_\alpha\partial_\beta h^{\alpha\beta})\right)\,,
\label{eq:epsilondef}
\end{equation}
 whereas $a_1=-1/2$, $a_{2,3}$ are two arbitrary
constants related to the model parameters $\alpha_{3,4}$,
and the tensors  $X^{(1,2,3)}_{\mu\nu}$ denote the interactions with the helicity-0 mode \cite{deRham:2010ik}
\begin{eqnarray}
  &&X_{\mu\nu}^{(1)}=
  -{1\over 2} \varepsilon_\mu^{~\alpha\rho\sigma}\varepsilon_{\nu~\rho\sigma}^{~\beta}\Pi_{\alpha\beta},\\
  &&X_{\mu\nu}^{(2)}=
  -{1\over 2} \varepsilon_\mu^{~\alpha\gamma\rho}\varepsilon_{\nu~~\rho}^{~\beta\delta}\Pi_{\alpha\beta}\Pi_{\gamma\delta},\\
  &&X_{\mu\nu}^{(3)}=\varepsilon_\mu^{~\alpha\gamma\rho}\varepsilon_{\nu}^{~\beta\delta\sigma}\Pi_{\alpha\beta}\Pi_{\gamma\delta}\Pi_{\rho\sigma}.
\end{eqnarray}
Here we defined $\Pi_{\mu\nu}\equiv \partial_\mu\partial_\nu\pi$, and 
$\Lambda_3$ represents the strong coupling scale of this theory.
One can easily check that 
this Lagrangian possess the diffeomorphism invariance, 
$x^\mu \to x^\mu + \xi^\mu$, 
and the Galileon symmetry, $\partial_\mu \pi \to \partial_\mu \pi + c_\mu $.
The structure of $X_{\mu\nu}^{(1,2,3)}$ is the same as the Galileon theory,
which ensures that the equation of motion remains second-order differential 
equation (i.e., this theory is free of Boulware-Deser ghost) and which also guaranties the existence of non-renormalization theorem \cite{deRham:2012ew}.

\subsection{Proxy theory from the decoupling limit}
We now want to covariantize the decoupling limit theory. 
The decoupling limit theory is only valid within the compton wavelength 
of massive graviton\footnote{
In order to explain the current accelerated expansion of the universe 
driven by the mass of graviton, 
it has to be order of the present Hubble horizon, $H_0$.}.
Once we covariantize the decoupling limit theory, 
the proxy theory is no longer massive gravity;
however, they share the same decoupling limit.
It would be very interesting to study the cosmology of the proxy theory
in order to see the differences to the original massive gravity theory.
After covariantizing the decoupling limit interactions the resulting interactions become \cite{deRham:2011by}
\begin{eqnarray}
h^{\mu\nu}X^{(1)}_{\mu\nu} &\longleftrightarrow& 
{1 \over 2} \sqrt{-g}\, 
\pi \varepsilon^{\mu\nu\rho\sigma}\varepsilon^{\alpha\beta}_{~~~\rho\sigma}
  R_{\mu\alpha\nu\beta} 
=-\sqrt{-g}\, \pi R,\\
h^{\mu\nu}X^{(2)}_{\mu\nu} &\longleftrightarrow& 
-{1 \over 2} \sqrt{-g}\, \, 
\varepsilon^{\mu\nu\rho\sigma}\varepsilon^{\alpha\beta\gamma}_{~~~~\sigma}
  R_{\mu\alpha\nu\beta} \partial_\rho \pi \partial_\gamma \pi
=-\sqrt{-g}\, \partial_\mu\pi\partial_\nu\pi G^{\mu\nu},\\
h^{\mu\nu}X^{(3)}_{\mu\nu}  &\longleftrightarrow& 
\sqrt{-g}\,\varepsilon^{\mu\nu\rho\sigma}\varepsilon^{\alpha\beta\gamma\delta}
R_{\mu\alpha\nu\beta} \partial_\rho \pi \partial_\gamma \pi \Pi_{\sigma\delta}
=-\sqrt{-g}\, \partial_\mu \pi \partial_\nu \pi \Pi_{\alpha\beta} L^{\mu\alpha\nu\beta}.
\end{eqnarray}
Here we used the fact that 
\begin{eqnarray}
\biggl[\sqrt{-g}
  \varepsilon^{\mu\nu\rho\sigma}\varepsilon^{\alpha\beta\gamma\delta}
  R_{\mu\alpha\nu\beta}\biggr]_{h}
=-\varepsilon^{\mu\nu\rho\sigma}\varepsilon^{\alpha\beta\gamma\delta}
  \partial_{\mu}\partial_{\alpha}\,h_{\nu\beta},
\end{eqnarray}
and the tensors $G_{\mu\nu}$ and $L^{\mu\alpha\nu\beta}$ are the Einstein and the dual Riemann tensors respectively,
\begin{eqnarray}\label{Einsteindual}
G^{\mu\nu}&=&R_{\mu\nu}-\frac12 R g_{\mu\nu},\\
L^{\mu\alpha\nu\beta}&=&2R^{\mu\alpha\nu\beta}+2(R^{\mu\beta}g^{\nu\alpha}+R^{\nu\alpha}g^{\mu\beta}-R^{\mu\nu}g^{\alpha\beta}-R^{\alpha\beta}g^{\mu\nu})\nonumber\\
&&+R(g^{\mu\nu}g^{\alpha\beta}-g^{\mu\beta}g^{\nu\alpha})\,.
\end{eqnarray}
Thus, the covariantization of the decoupling limit Lagrangian \eqref{lagr} gives birth to the following proxy theory
\begin{equation} \label{eq:TotalAction}
\mathcal{L}=\sqrt{-g}\left({M_{\rm Pl}^2 \over 2} R+\mathcal{L}^\pi(\pi,g_{\mu \nu})+\mathcal L^{\rm matter}(\psi,g_{\mu \nu})\right)\,,
\end{equation}
where the Lagrangian for $\pi$ is
\begin{equation} \label{eq:covJor}
\mathcal{L}^\pi=M_{\rm Pl} \left(-\pi R-\frac{a_2}{\Lambda^3}\partial_\mu\pi\partial_\nu\pi G^{\mu\nu}-\frac{a_3}{\Lambda^6}\partial_\mu\pi\partial_\nu\pi \Pi_{\alpha\beta} L^{\mu\alpha\nu\beta}\right)\,.
\end{equation}
These correspondences relate the decoupling limit of massive gravity 
to the subclass of Horndeski scalar-tensor interactions. This proxy theory represents a theory of GR on top of which a new scalar degree of freedom is added, which is non-minimally coupled to gravity \footnote{See also in \cite{Charmousis:2011bf} where similar interactions were considered, even though unrelated to massive gravity.}.
The Galileon symmetry is broken by covariantizing 
the decoupling limit Lagrangian as in the most general second order 
scalar-tensor theory.
Furthermore the constant shift symmetry, $\pi \to \pi + c$, is not even preserved
by covariantization. 
Note that $\pi R$ term satisfies the constant shift symmetry 
at linear level; however,
the nonlinear corrections in $\pi R$ term break the shift symmetry.

\subsection{Proxy theory as a subclass of Horndeski scalar-tensor theories}
As mentioned above, the proxy theory is a subclass of Horndeski scalar-tensor theories which describes the most general scalar tensor interactions with second order equations of motion. The general functions of the Horndeski interactions can be related with the proxy theory. The Horndeski action is given by the following action
\begin{eqnarray}
S=\int d^4x\sqrt{-g}\left(\sum_{i=2}^5\mathcal{L}_i+\mathcal{L}_m\right),
\end{eqnarray}
with
\begin{eqnarray}\label{HorndeskiGal}
\mathcal{L}_2&=&K(\pi,X)\nonumber\\
\mathcal{L}_3&=&-G_3(\pi,X)[\Pi]\nonumber\\
\mathcal{L}_4&=&G_4(\pi,X)R+G_{4,X}\left([\Pi]^2-[\Pi^2]\right)\nonumber\\
\mathcal{L}_5&=&G_5(\pi,X)G_{\mu\nu}\Pi^{\mu\nu}-\frac16G_{5,X}\left([\Pi]^3-3[\Pi][\Pi^2]+2[\Pi^3]\right) ,
\end{eqnarray}
where the arbitrary functions $K$, $G_3$, $G_4$ and $G_5$ depend on the scalar field $\pi$ and its derivatives $X=-\frac12(\partial\pi)^2$ and furthermore $G_{i,X}=\partial G_i/\partial X$ and $G_{i,\pi}=\partial G_i/\partial \pi$. The proxy theory corresponds to the case for which the above functions take the concrete following forms \cite{Kimura:2011qn, Kobayashi:2010aa}
\begin{eqnarray}\label{proxy_versus_horndeski}
K(\pi,X)&=&0\nonumber\\
G_3(\pi,X)&=&0\nonumber\\
G_4(\pi,X)&=&\frac{M_{\rm Pl}^2}{2}-M_{\rm Pl}\pi-\frac{M_{\rm Pl}}{\Lambda^3}a_2X\nonumber\\
G_5(\pi,X)&=&3\frac{M_{\rm Pl}}{\Lambda^6}a_3X.
\end{eqnarray}
The Horndeski scalar-tensor theories represent an interesting class of modified gravity models. However, with the general functions $K$, $G_3$, $G_4$ and $G_5$ it is hard to study the entire class at once. In the literature, there has been some attempts of parametrizing the theory in a way that would allow to investigate the theory as a whole in order to favor or rule out by observations. \cite{Amendola:2012ky, Gomes:2013ema}. The interesting point in the proxy theory is that it has its original motivation in massive gravity and has to have the explicit form constructed out of the decoupling limit. Thus, this construction relates two important classes of modified gravity theories, namely massive gravity and Horndeski theories.

\section{dynamical system analysis}
\label{sec:dyn_analy}

\subsection{Field equations}
From now on, we discuss the properties of cosmological solutions. 
We first work on the spatially flat Friedmann-Robertson-Walker metric,
$ds^{2}=-dt^{2}+a^{2}(t)\delta_{ij}dx^{i}dx^{j}$, 
and then the gravity equations are given by
\begin{eqnarray}
3M_{\mathrm{Pl}}^2H^2&=&\rho_{\pi}+\rho_m,
\label{MFE}
\\
-M_{\mathrm{Pl}}^2\left(2\dot{H}+3H^2\right)&=&p_{\pi},
\label{MFE2}
\end{eqnarray}
where $H(=\dot a/a)$ is the Hubble parameter, $\rho_m$ 
is the energy density of matter, and
the energy density and the pressure of the Galileon field are defined by
\begin{eqnarray}
  \rho_{\pi}&=&
\mpl \biggl(6H^2\pi+6H\dot\pi-\frac{9a_2}{\Lambda^3}H^2\dot\pi^2
-\frac{30a_3}{\Lambda^6}H^3\dot\pi^3
\biggr),
\\
  p_{\pi}&=&\mpl \biggl(
-2(2\dot{H}+3H^2)\pi-4 H {\dot \pi }-2{\ddot \pi}
+\frac{a_2}{\Lambda^3}(3H^2 {\dot \pi}^2 + 2 {\dot H}{\dot \pi}^2+4H{\dot \pi}{\ddot \pi})\nonumber\\
&&~~~~~~~~+\frac{6a_3}{\Lambda^6}(2H^3{\dot \pi}^3+2H {\dot H} {\dot \pi}^3+3H^2{\dot \pi}^2 {\ddot \pi})
\biggr),
\end{eqnarray}
and the equation of motion for $\pi$ in the FRW space-time is given by
\begin{eqnarray}
\frac{6a_2}{\Lambda^3}\biggl(3H^3\dot\pi+2H\dot H\dot\pi+H^2\ddot\pi\biggr)
+\frac{18a_3}{\Lambda^6}\biggl(3H^2\dot H\dot\pi^2+3H^4\dot\pi^2+2H^3\dot\pi\ddot\pi\biggr)
={\bar R},
\end{eqnarray}
where ${\bar R}$ is the Ricci scalar evaluated in FRW metric, ${\bar R}=6({\dot H} + 2H^2)$. 
This field equation for $\pi$ can be recast in a compact form,
\begin{eqnarray}
  \ddot\phi+3H\dot\phi-{\bar R}=0,
\label{fieldeq}
\end{eqnarray}
where the new field $\phi$ is defined by 
\begin{eqnarray}
  \dot\phi=H^2\biggl(\frac{6a_2}{\Lambda^3}\dot\pi
  +\frac{18a_3}{\Lambda^6}\dot\pi^2H
\biggr)\,.
\label{defphi}
\end{eqnarray}

\subsection{de Sitter regime}

The de Sitter solutions which had been found in  \cite{deRham:2011by} are only valid in the approximation $H\pi \ll \dot \pi$. In  \cite{deRham:2011by}, it has been shown that de Sitter is a legitimate solution when such an approximation holds. In the following we will study the validity of this approximation in more detail. 
In a pure de Sitter background with constant expansion rate $H_{dS}$, the exact homogeneous field equation reads
\begin{equation}
\frac{6H_{dS}^2}{\Lambda^3}\left(a_2+6a_3\frac{H_{dS}}{\Lambda^3}\dot{\pi}\right)\ddot{\pi}+18\frac{H_{dS}^3}{\Lambda^3}\left(a_2+3a_3\frac{H_{dS}}{\Lambda^3}\dot{\pi}\right)\dot{\pi}=12H_{dS}^2.
\end{equation}
In \cite{deRham:2011by}, this equation together with Friedman equation were solved by using the approximation $\pi H\ll \dot{\pi}$ and the Ansatz of constant $\dot{\pi}$. However, this equation can actually be exactly solved without making such an approximation and the corresponding solution exhibits the two following branches for $\dot{\pi}$:
\begin{equation}
\dot{\pi}=\frac{-a_2\Lambda^3\pm e^{-\frac32H_{dS}t}\sqrt{4a_3 e^{C_1}+(a_2^2+8a_3)e^{3H_{dS} t}\Lambda^6}}{6a_3 H_{dS}},
\end{equation}
with $C_1$ an integration constant. At late times, one can easily see that $\dot{\pi}$ evolves towards the constant value
\begin{equation}
\dot{\pi}(t\gg H_{dS}^{-1})\simeq -\frac{\Lambda^3}{6a_3 H_{dS}}\left[a_2\pm\sqrt{a_2^2+8a_3}  \right].
\end{equation}
This coincides with the finding in \cite{deRham:2011by} when assuming the Ansatz $\dot{\pi}=q \Lambda^3 / H_{dS}$, showing that such a solution is indeed the attractor solution in a de Sitter background. It is important to notice that this solution has been obtained by assuming that the de Sitter background is not driven by the $\pi$ field, but by some other independent effective cosmological constant. Now we want to study if such an effective cosmological constant can be generated by the $\pi$ field itself so that de Sitter is an actual solution of the system. From the above solution for $\dot{\pi}$,  it is straightforward to obtain the solution for $\pi$ by means of a simple integration
\begin{equation}
\pi(t\gg H_{dS}^{-1})\simeq -\frac{\Lambda^3}{6a_3 H_{dS}}\left[a_2\pm\sqrt{a_2^2+8a_3}  \right]t+C_2 ,
\end{equation}
where $C_2$ is another integration constant. If we plug this solution into the energy density of $\pi$ (which gives the r.h.s. of Friedman equation), we obtain
\begin{align}
\rho_\pi\simeq&\frac{M_p\Lambda^3}{18}\left[108C_2\frac{H_{dS}^2}{\Lambda^3}+\left(\frac{a_2^3}{a_3^2}+6\frac{a_2}{a_3}\right)\pm\left(\frac{a_2^2+2a_3}{a_3^2}\right)\sqrt{a_2^2+8a_3} \right] \nonumber\\
&-\frac{M_p\Lambda^3}{a_3}\left(a_2\pm\sqrt{a_2^2+8a_3}\right)H_{dS}t.
\end{align}
At early times when $H_{dS}t\ll1$ we can neglect the second term in this expression, the energy density of the $\pi$ field is approximately constant, as it corresponds to a de Sitter solution. However, we must keep in mind that this solution is actually valid at late times and, in that case, the second term growing linearly with time drives the energy density evolution and, thus, de Sitter cannot be the solution. This also agrees with the fact that the condition $\pi H\ll\dot{\pi}$ will be eventually violated at late times because the scalar field grows in time, whereas $H$ and $\dot{\pi}$ are assumed to be constant. One might think that a way out would be to tune the parameters so that $a_2\pm\sqrt{a_2^2+8a_3}=0$. However, the only solution to this equation is $a_3=0$, which represents a singular value. In fact, if we take the limit $a_3\rightarrow0$ in the above solution, we obtain $\rho_\pi\rightarrow 6C_2H_{dS}M_p+4M_p\Lambda^3H_{dS}t/a_2$ so the growing term remains. From this simple analysis, it seems that de Sitter cannot exist as an attractor solution of the phase map, but it can only represent transient regimes. This can in turn be useful for inflationary models where the accelerated expansion needs to end, but it is less appealing as dark energy model.

\subsection{Phase analysis without matter component}

In the following, we will make this simple analysis more rigorous and look at it in more detail. In order to obtain a general overview of the class of cosmological solutions that one can expect to find in the proxy theory, we shall perform a dynamical system analysis. This will give us the critical points of the cosmological equations as well as their stability. The first step to perform the dynamical system analysis will be to obtain the equations to be analyzed. Since we are interested in cosmological solutions, the metric will be assumed to take the FLRW form with flat spatial sections.
The most convenient time variable for the analysis will be the number of e-folds $N\equiv\ln a$. The equation of motion for the $\pi$ field in terms of this time variable is given by
\begin{align}
&\left(a_2+6a_3H^2\frac{\pi'}{\Lambda^3}\right)\pi''+3\left[a_2\left(1+\frac{H'}{H}\right)+\frac{a_3H^2}{\Lambda^3}\left(3+5\frac{H'}{H}\right)\pi'\right]\pi'\nonumber\\
&=2\frac{\Lambda^3}{H^2}\left(1+\frac{H'}{2H}\right),
\end{align}
where the prime denotes derivative with respect to $N$. In addition to this equation, we also need the corresponding Einstein equations, which in our case are given by
\begin{eqnarray}
H^2&=&\frac{1}{6M_p^2}\rho_\pi\\
2HH'+3H^2&=&-\frac{1}{2M_P^2}p_\pi,
\end{eqnarray}
where we have used that $dN=Hdt$ and $\rho_\pi$ and $p_\pi$ are the energy density and pressure of the $\pi$ field expressed in terms of $N$. We have now three equations for the two variables $\pi$ and $H$. Of course, not all three equations are independent. In order to reduce these equations to the form of an autonomous system, we will first use the Friedman constraint to obtain an expression for $\pi$ in terms of $\pi'$ and $H$. The resulting expression will constitute a constraint for $\pi$ and will allow us to get rid of its dependence in the remaining equations so that we end up with dependence only on $H$, $H'$, $\pi'$ and $\pi''$. This will result very useful since it reduces the number of variables in our autonomous system. In fact, we can use $y\equiv\pi'$ as one of our dynamical variables and, then, we have a system of two first order differential equations for $y$ and $H$. After some simple algebra, one can reduce the equations to the following autonomous system:
\begin{eqnarray}
\frac{dy}{dN}=-\frac{1+3b_2H^2y+(25b_3-9b_2^2)H^4y^2-87b_2b_3H^6y^3-180b_3^2H^8y^4}{1-6b_2H^2y+6(b_2^2-5b_3)H^4y^2+52b_2b_3H^6y^3+105b_3^2H^8y^4}y\nonumber\\\nonumber\\
\frac{dH}{dN}=-\frac{2-8b_2H^2y+(9b_2^2-33b_3)H^4y^2+72b_2b_3H^6y^3+135b_3^2H^8y^4}{1-6b_2H^2y+6(b_2^2-5b_3)H^4y^2+52b_2b_3H^6y^3+105b_3^2H^8y^4}H,\nonumber\\\:
\label{autonomous}
\end{eqnarray}
where we have introduced the rescaled parameters $b_2\equiv a_2 M_p^3 / \Lambda^3$ and $b_3\equiv a_3 M_p^6 / \Lambda^6$. One can immediately see that $H=y=0$ is a stable critical point which is independent of the parameters and corresponds to the vacuum Minkowski solution. For the remaining critical points, we need to solve the equations
\begin{eqnarray}
1+3b_2H^2y+(25b_3-9b_2^2)H^4y^2-87b_2b_3H^6y^3-180b_3^2H^8y^4&=&0,\nonumber\\
2-8b_2H^2y+(9b_2^2-33b_3)H^4y^2+72b_2b_3H^6y^3+135b_3^2H^8y^4&=&0.\nonumber
\end{eqnarray}
To solve these equations, it will be convenient to introduce a new rescaling as $\hat{y}\equiv H^2y b_2$ and the new constant $c_3\equiv b_3/b_2^2=a_3/a_2^2$. Then, the previous equations can be written in the simpler form
\begin{eqnarray}
1+3\hat{y}+(25c_3-9)\hat{y}^2-87c_3\hat{y}^3-180c_3^2\hat{y}^4&=&0,\\
2-8\hat{y}+(9-33c_3)\hat{y}^2+72c_3\hat{y}^3+135c_3^2\hat{y}^4&=&0.
\end{eqnarray}
As we can see, we have an overdetermined system of equations so that solutions cannot be found for arbitrary $c_3$. In fact, the above equations can be solved for $\hat{y}$ and $c_3$ in order to obtain the models with additional critical points. Remarkably, there is only one real solution for these equations and is given by $c_3\simeq0.094$ and $\hat{y}\simeq-3.99$. Notice that this in fact does not represent one single critical point for the autonomous system, but a curve of critical points in the plane $(y,H)$. The obtained result implies that pure de Sitter does not correspond to a critical point of the proxy theory and can only exist as a transient regime, as we had anticipated from our previous simple analysis.

Another interesting feature of the autonomous system is the existence of separatrices in the phase map determined by the curve along which the denominators in (\ref{autonomous}) vanish, i.e.
\begin{equation}
1-6b_2H^2y+6(b_2^2-5b_3)H^4y^2+52b_2b_3H^6y^3+105b_3^2H^8y^4=0.
\end{equation}
This curve can be simplified if we use our previously defined rescaled variable $\hat{y}$ and parameter $c_3$, in terms of which the separatrix is determined by
\begin{equation}
1-6\hat{y}+6(1-5c_3)\hat{y}^2+52c_3\hat{y}^3+105c_3^2\hat{y}^4=0,
\label{separatrix}
\end{equation}
which is a quartic polynomial equation. Being the independent term and the highest power coefficient both positive, this equation does not always have real solutions so the separatrix does not exist for arbitrary parameters. Indeed, the previous equation determines a curve in the plane $(\hat{y},c_3)$, which can be regarded as the function
\begin{equation}
c_3=\frac{15-26\hat{y}\pm\sqrt{2}\sqrt{60-75\hat{y}+23\hat{y}^2}}{105\hat{y}^2}.
\label{eqseparatrix}
\end{equation}
This function has been plotted in Fig. \ref{separatrix}. As we can see in that figure, the value of $c_3$ determines the number of real solutions and, therefore, the number of separatrices in the phase map of the autonomous system. We find that for $c_3>0$, the system always exhibits 4 separatrices. When $c_3=0$, the cubic and quartic terms of the separatrix equation vanish, so we only have two real solutions. In the cases with $0>c_3>-0.093$, the system has 4 separatrices again. When $-0.093>c_3>-0.215$, there are only 2 separatrices and, finally, for $c_3<-0.215$, the equation has no real solutions and, therefore, it does not generate any separatrix. Special cases are $c_3=-0.093$ with 3 separatrices and $c_3=-0.215$ with only one separatrix. All this can be clearly seen in Fig. \ref{separatrix}.

\begin{figure}
\includegraphics[width=15.cm]{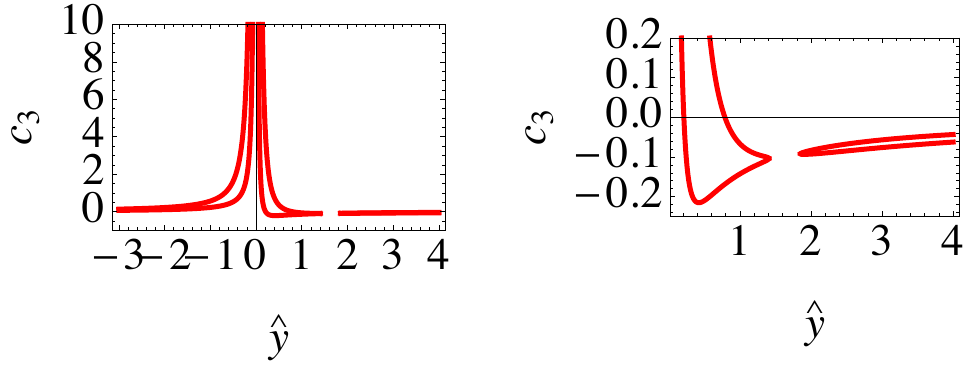}
\caption{In this plot we show the curve determined by  Eq. (\ref{separatrix}) in the plane $(\hat{y}=H^2\pi'b_2,c_3=a_3/a_2^2)$. The right panel shows a detail to see more clearly the structure of the corresponding area. As explained in the main text, the value of the parameter $c_3$ determines the number of separatrices in the phase map. }
\label{separatrix}
\end{figure}

\begin{figure}[h!]
\includegraphics[width=8.cm]{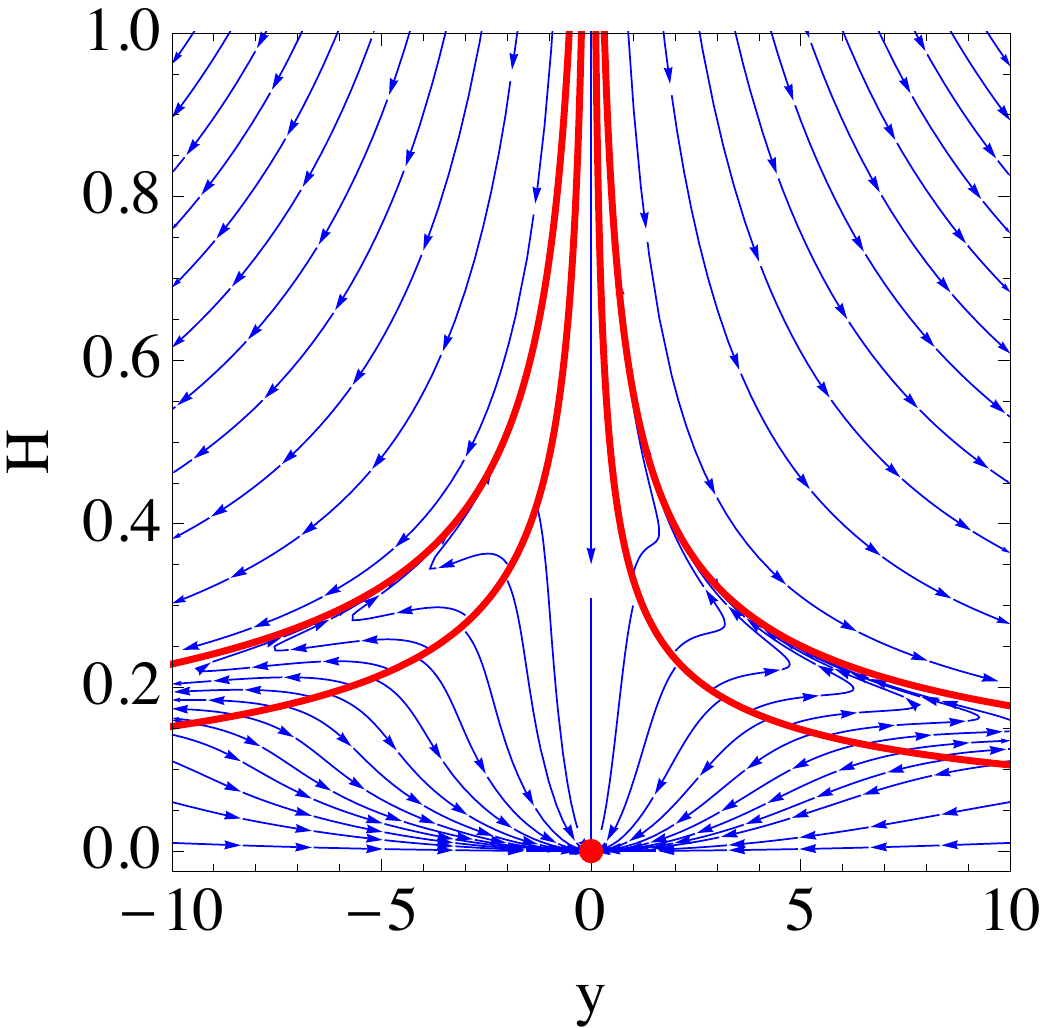}
\includegraphics[width=8.cm]{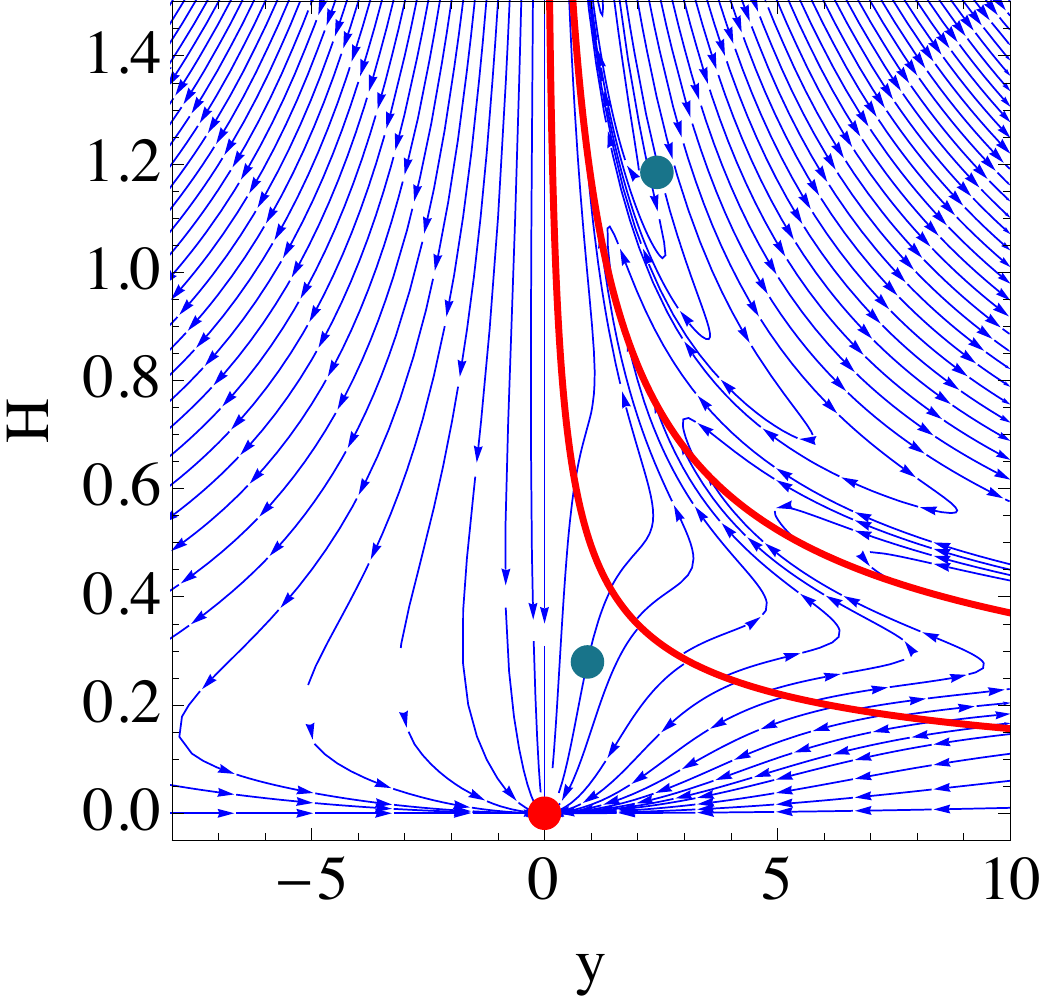}
\caption{In this figure we show two examples of phase map portraits of the dynamical autonomous system for $b_2=1$ and $b_3=1.5$ (with $c_3=1.5$) in the left panel and  $b_2=1$ and $b_3=-0.1$ (with $c_3=-0.1$) in the right panel. These values have been chosen to show examples with $c_3>0$ (always with 4 separatrices) and $c_3<0$ with 2 separatrices (see main text and Fig. \ref{separatrix}) The red lines represent the corresponding separatrices and the red point denotes the Minkowski vacuum solution. We can see that this solution is indeed an attractor. Concerning the attracting behaviour of the separatrices, we can see that the upper ones behave as attractors, whereas the lower ones act as repelers. In the right panel, we additionally indicate with green points the analytical solutions found in \cite{deRham:2011by} under the approximation $\pi H\ll\dot{\pi}$. }
\label{phasemap}
\end{figure}

 If the solutions of Eq. (\ref{eqseparatrix}) are denoted by $\hat{y}=y^*_i$, then, the separatrices are given by the curves $y=b_2y^*_i/H^2$ or, equivalently, $H=\pm\sqrt{b_2y^*_i/y}$ in the phase map. Notice that, depending on the sign of  $b_2y_i^*$, the corresponding separatrix will only exist in the semi-plane $y>0$  or $y<0$ for $b_2y_i^*>0$ or $b_2y_i^*<0$ respectively. This can be seen in the examples shown in Fig. \ref{phasemap} where we have plotted the phase maps corresponding to two characteristic cases, namely, one with $c_3=1.5$ (which has 4 separatrices and positive $c_3$) and one with $c_3=-0.1$ (which has only 2 separatrices and negative $c_3$). One interesting feature that we can observe in both cases is the attracting nature of the upper separatrices, whereas the lower ones behave as repellers. Remarkably, the attracting separatrices do not behave as asymptotic attractors, but the trajectories actually hit the separatrix and the universe encounters a singularity.
 
The phase map shown in the right panel corresponds to parameters satisfying all the existence and stability requirements obtained in \cite{deRham:2011by} from the approximate analytical solutions. The green points in the phase map denote the solutions that had been identified in \cite{deRham:2011by} with stable self-accelerating solutions. However, we can see now that the eventual attractor solution is not actually de Sitter but the Minkowksi vacuum solution. The stability condition for such a solution actually corresponds to the convergence of the nearby trajectories.

\begin{figure}[h!]
\includegraphics[width=8.cm]{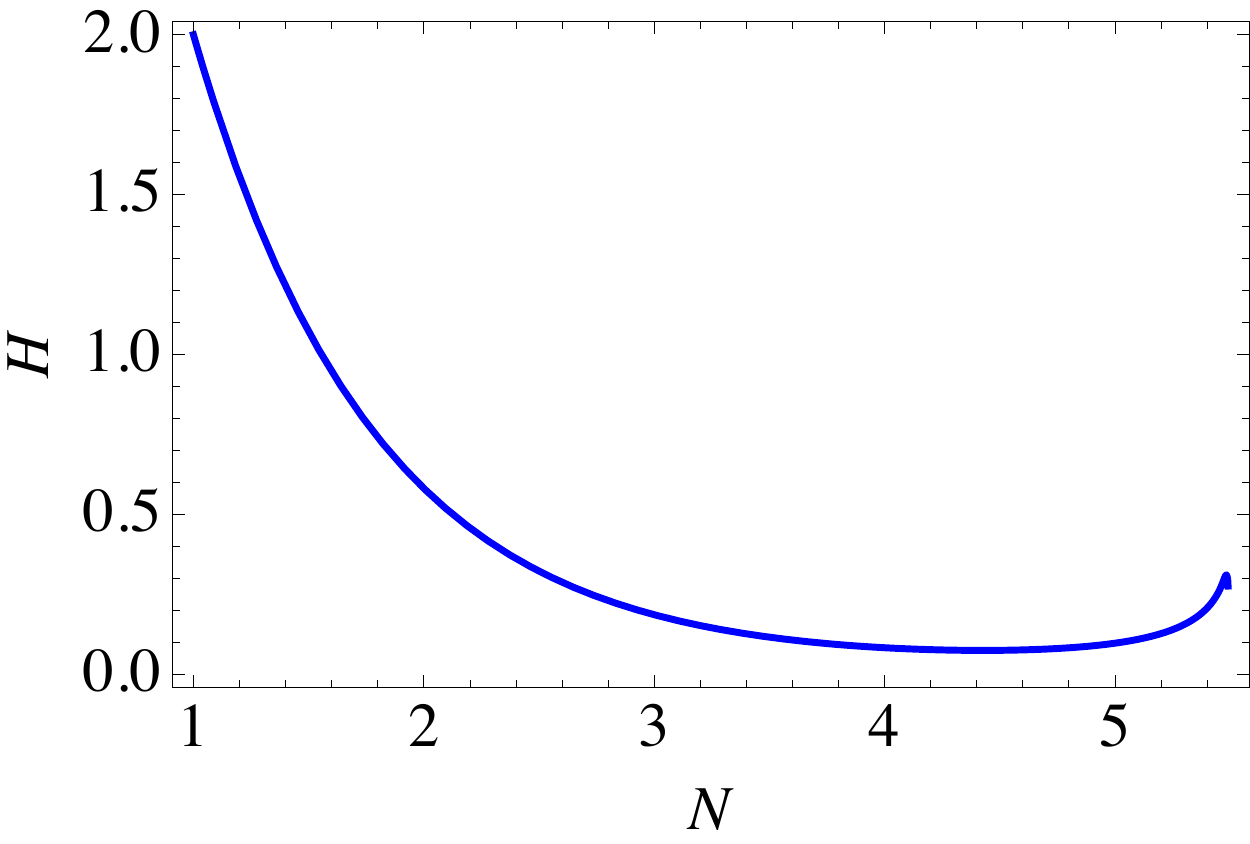}
\includegraphics[width=8.cm]{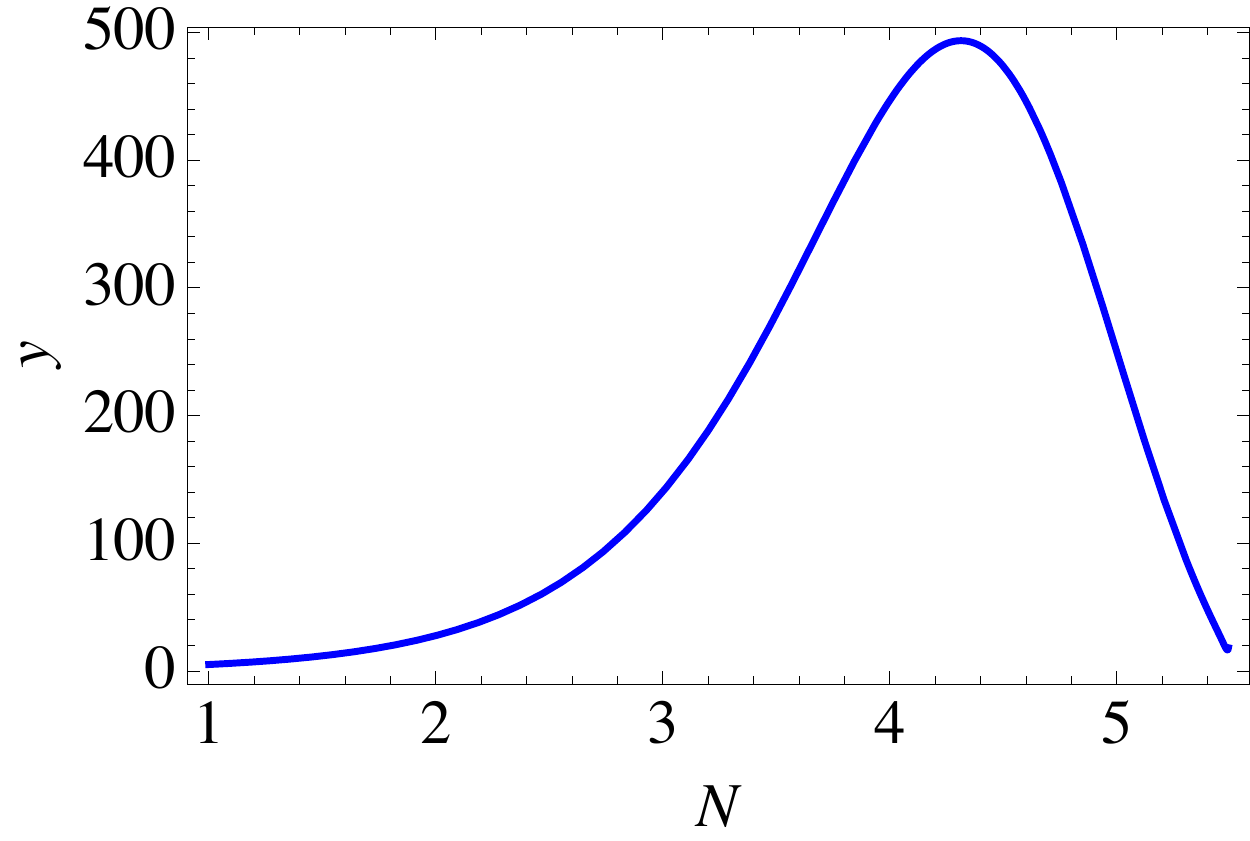}
\caption{In this figure we show the numerical solution for $H$ (left panel) and $y$ (right panel) with the initial conditions $H_{ini}=2$ and $y_{ini}=5$. We can see the transient period of quasi de Sitter expansion in the evolution of $H$ corresponding to the turnover and how it lasts for barely $1-2$ e-folds. In addition, we can see the discussed singularity corresponding to the moment when the trajectory reaches the separatrix at a finite number of e-folds. }
\label{numsolution}
\end{figure}

It is worthwhile pointing out once more that, although (quasi) de Sitter solutions do not exist as critical points in the phase maps, it is possible to have transient regimes with quasi de Sitter expansion. One possibility where such transient regimes can be found correspond to the trajectories above the upper separatrix in the right panel of Fig. \ref{phasemap}. These trajectories initially evolve towards large values of $y$, but, at some point, there is a turnover where it goes towards smaller values of $y$. While this turnover is taking place, the value of $H$ can remain nearly constant for some time and, thus, we can have a period of quasi de Sitter expansion. The number of e-folds corresponding to this transient regime depends on the parameters and the initial conditions, but it is generally quite small (see Fig. \ref{numsolution} where we plot the evolution of one particular solution).

In order to study the properties of the dynamical system near the separatrix, we will rewrite the autonomous system in terms of the variable $\hat{y}$, since, as suggested from our previous analysis, the equations will look simpler. In particular, the separatrices will become  straight vertical lines in this variable and the behaviour of the trajectories near them can be straightforwardly studied. In such variables, the autonomous system reads
\begin{eqnarray}
\frac{d\hat{y}}{dN}=-\frac
{5-13\hat{y}+(9-41c_3)\hat{y}+57c_3\hat{y}^3+90c_3^2\hat{y}^4}
{1-6\hat{y}+6(1-5c_3)\hat{y}^2+52c_3\hat{y}^3+105c_3^2\hat{y}^4}\hat{y}\nonumber\\\nonumber\\
\frac{dH}{dN}=-\frac
{2-8\hat{y}+(9-33c_3)\hat{y}^2+72c_3\hat{y}^3+135c_3^2\hat{y}^4}
{1-6\hat{y}+6(1-5c_3)\hat{y}^2+52c_3\hat{y}^3+105c_3^2\hat{y}^4}H.
\label{autonomous2}
\end{eqnarray}
As we anticipated, the equations look simpler in these variables. In particular, the equation for $\hat{y}$ completely decouples from the equation for the Hubble expansion rate. Near the separatrix located at $y_s$ we can expand $\hat{y}=\hat{y}_s+\delta\hat{y}$ and obtain the leading terms of the above equations, given by
\begin{eqnarray}
\frac{d\delta\hat{y}}{dN}=\frac{k_y}{\delta\hat{y}},\quad\quad
\frac{dH}{dN}=\frac
{k_H}
{\delta\hat{y}}H,
\label{autonomous2}
\end{eqnarray}
with
\begin{eqnarray}
k_y&\equiv&-\frac
{5-13\hat{y}_s+(9-41c_3)\hat{y}_s+57c_3\hat{y}_s^3+90c_3^2\hat{y}_s^4}
{1-6+12(1-5c_3)\hat{y}_s+156c_3\hat{y}^2_s+420c_3^2\hat{y}_s}\hat{y}_s\\\nonumber\\
k_H&\equiv&-\frac{2-8\hat{y}_s+(9-33c_3)\hat{y}_s^2+72c_3\hat{y}_s^3+135c_3^2\hat{y}_s^4}
{1-6+12(1-5c_3)\hat{y}_s+156c_3\hat{y}^2_s+420c_3^2\hat{y}_s}.
\end{eqnarray}
Now, it is straightforward to read the conditions for the separatrix to attract the trajectories. Notice that the attracting or repelling nature of the separatrix will be the same from both sides of it. Thus, whenever $k_y$ is negative, the separatrix will represent an attractor of the phase map, whereas it will be a repeller for positive $k_y$. 

The equation for $\delta\hat{y}$ near the separatrix can be easily integrated to give
\begin{equation}
\delta\hat{y}(N)\simeq\pm\sqrt{2k_yN+C_y},
\end{equation}
with $C_y$ an integration constant and the two branches correspond to both sides of the seperatrix. If the separatrix is an attractor, we have that $k_y$ is negative and, therefore, the solution only exists until $N_s=-\frac{C_y}{2k_y}$, confirming our previous statement that the trajectories do not approach asymptotically the separatrix, but they hit it and end there. On the other hand, with the solution for $\delta\hat{y}$, we can also obtain the solution for $H$, which is given by
\begin{equation}
H(N)=C_He^{\pm\frac{k_H}{k_y}\sqrt{2k_yN+C_y}},
\end{equation}
with $C_H$ another integration constant. We see that the Hubble expansion rate does not diverge at the separatrix, but it goes to the constant value $C_H$ so that the energy density of the field remains finite. However, the derivative of the Hubble expansion rate near the separatrix evolves as
\begin{equation}
 \dot{H}\simeq H^2\frac{k_H}{\sqrt{2k_yN+C_y}},
 \end{equation}
 so it goes to infinity as it approaches the separatrix. This signals a divergence in the pressure of the scalar field when the trajectory hits the separatrix so we find a future sudden singularity. This kind of singularity was first studied in \cite{Barrow:2004xh} and corresponds to the type II according to the classification performed in \cite{Nojiri:2005sx}.

\subsection{Phase analysis with matter component}

So far, in our study we have focused on the case when only the $\pi$ field contributes to the energy density of the universe and we have neglected any other possible component that might be present. We have shown that the only critical point is the pure vacuum Minkowski solution with $H=y=0$. Moreover, we have shown that the separatrices can also act as attractors of the phase map and, when this happens, the evolution ends in a singularity where the derivative of the Hubble expansion rate diverges. In order to have a more realistic scenario, at least a matter component should be included. This will add a new dimension to the phase space and, thus, a new phenomenology is expected to arise. In particular, it could change some stability requirements and additional critical points might appear. Therefore, let us discuss in the following the case with matter fields.\\
If we include a pressureless matter component and use the variables $H$, $\hat{y}$ and\footnote{Notice the factor $b_2$ in this definition of the matter density parameter that does not appear in the usual definition.} $\Omega_m\equiv\rho_mb_2/(6H^2)$, to describe the extended cosmological evolution, the corresponding autonomous system reads
\begin{eqnarray}
\frac{d\hat{y}}{dN}&=&-\frac
{\left(5-13\hat{y}+(9-41c_3)\hat{y}^2+57c_3\hat{y}^3+90c_3^2\hat{y}^4\right)\hat{y}+(1-3\hat{y}-9c_3\hat{y}^2)H^2\Omega_m}
{1-6\hat{y}+6(1-5c_3)\hat{y}^2+52c_3\hat{y}^3+105c_3^2\hat{y}^4-2(1+6c_3\hat{y})H^2\Omega_m}\nonumber\\\nonumber\\
\frac{dH}{dN}&=&-\frac
{2-8\hat{y}+(9-33c_3)\hat{y}^2+72c_3\hat{y}^3+135c_3^2\hat{y}^4-3(1+6c_3\hat{y})H^2\Omega_m}
{1-6\hat{y}+6(1-5c_3)\hat{y}^2+52c_3\hat{y}^3+105c_3^2\hat{y}^4-2(1+6c_3\hat{y})H^2\Omega_m}H\nonumber\\\nonumber\\
\frac{d\Omega_m}{dN}&=&\frac
{1+2\hat{y}+24c_3\hat{y}^2-12c_3\hat{y}^3-45c_3^2\hat{y}^4}
{1-6\hat{y}+6(1-5c_3)\hat{y}^2+52c_3\hat{y}^3+105c_3^2\hat{y}^4-2(1+6c_3\hat{y})H^2\Omega_m}\Omega_m.\nonumber\\
\label{autonomousmatter}
\end{eqnarray}
Since we are seeking for critical points with $\Omega_m\ne0$, 
we can solve for it from the vanishing of $d\hat{y}/dN$ to obtain the expression
\begin{equation}
\Omega_mH^2=\frac{5-13\hat{y}+(9-41c_3)\hat{y}^2+57c_3\hat{y}^3+90c_3^2\hat{y}^4}{-1+3\hat{y}+9c_3\hat{y}^2}\hat{y},
\end{equation}
for the potential new critical points. 
Then, we can plug this relation into the remaining two equations 
given by the vanishing of $dH/dN$ and $d\Omega_md/N$ to obtain the critical points. 
However, when doing so we end up with the solutions
\begin{equation}
c_3= \frac{4{\hat y}^2-2{\hat y}^3\pm \sqrt{21{\hat y}^4-6{\hat y}^5+4{\hat y}^6}}{15{\hat y}^4},
\end{equation}
which is incompatible for any value of $c_3$. 
Therefore, the inclusion of matter does not introduce new critical points in the phase map.

\subsection{Covariantization of the new kinetic interactions}
\begin{figure}[h!]
   \begin{center}
 \includegraphics[width=10.cm]{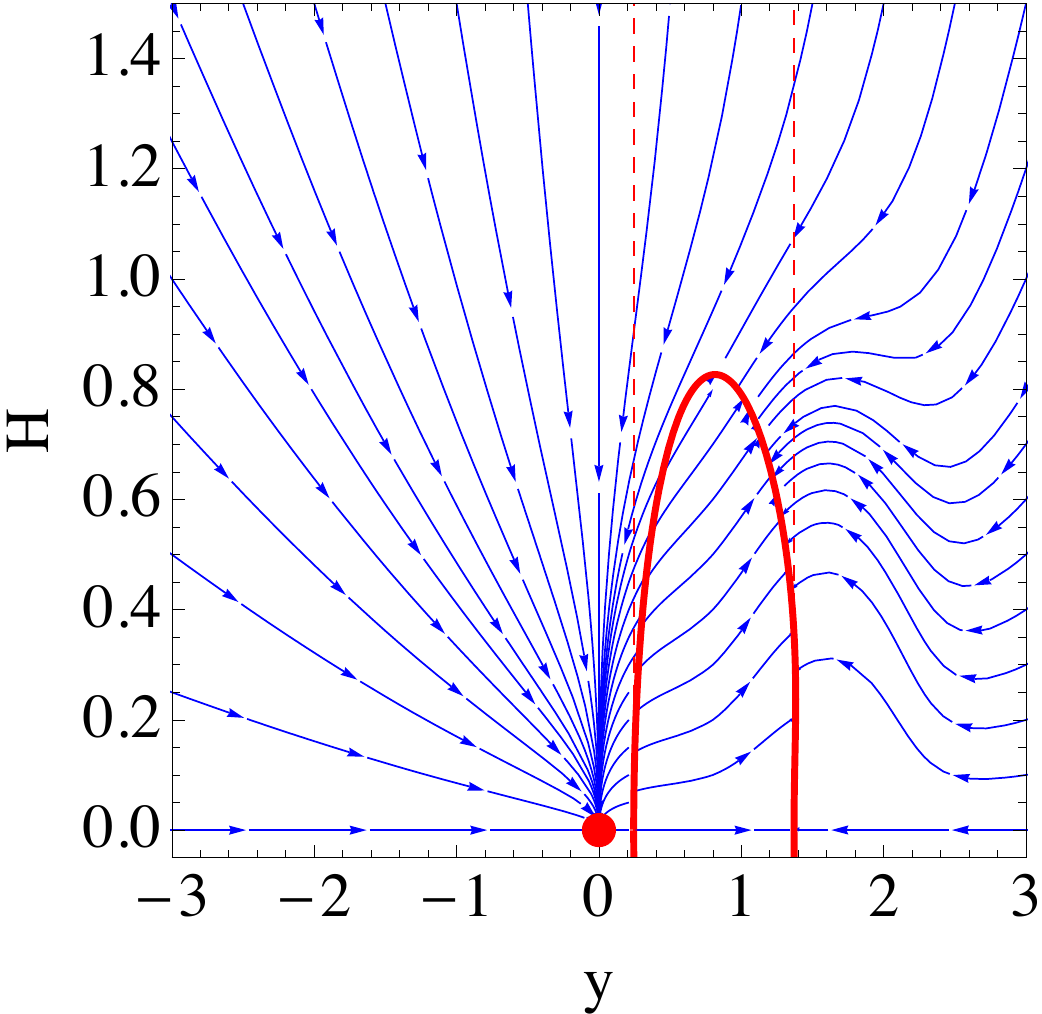}
 \caption{ In this figure we show an example of the phase map for the proxy theory with the additional Gauss-Bonnet term coming from the covariantization of the decoupling limit of the derivative interactions $\pi \mathcal L_{GM}$. One can clearly see that de Sitter solution is still not an attractor of the cosmological evolution leaving the Minkowski solution as the only existing attractor solution. Thus, the inclusion of this term does not change the cosmological properties of the proxy theory. The red line denotes the separatrix.}
 \label{phasemapGB}
  \end{center}
 \end{figure}

Above we have seen that the only critical point existing 
in the phase map of the proxy theory (even if we include a dust component) 
is the vacuum Minkowski solution. The proxy theory was constructed from the decoupling limit of the potential interactions of massive gravity. The mass and potential interactions of the graviton breaks the diffeomorphism invariance. Therefore one might wonder whether or not there exist derivative interactions for the graviton which break diffeomorphism invariance but still give rise to only five propagating physical degrees of freedom. In the literature exactly this question about the existence of new kinetic interactions was investigated \cite{deRham:2013tfa, Kimura:2013ika, Ohara:2014vua}. Possible terms of the form 
\begin{eqnarray}
&& {\cal K}_{\mu\nu}G^{\mu\nu} \nonumber\\
&& {\cal K}_{\mu\nu} {\cal K}_{\alpha\beta}L^{\mu\nu\alpha\beta},
\end{eqnarray}
have been considered and unfortunately shown that they contain ghost degree of freedom. Nevertheless, we can consider the decoupling limit of these interactions and covariantize them in a similar way as for the potential interactions. To first order in $h$ these interactions do not give any non-trivial interactions and are identically zero up to total derivatives. The second order interaction in $h$ of the interaction $K_{\mu\nu}G^{\mu\nu}$ gives rise to a ghost degree of freedom after covariantization and therefore we will not consider this contribution. On the other hand, from the interaction $K_{\mu\nu}K_{\alpha\beta}L^{\mu\nu\alpha\beta}$ the only second order contribution in $h$ which gives rise to ghost-free interaction is  ${\cal L}_{\rm DI}=\varepsilon^{\mu\nu\rho\sigma}\varepsilon^{\alpha\beta\gamma\delta}\partial_\mu\partial_\alpha h_{\nu\beta} h_{\rho\gamma} \partial_\sigma \partial_\delta \pi$ \cite{Folkerts:2011ev, Hinterbichler:2013aa}.
Covariantization of this decoupling limit Lagrangian ${\cal L}_{\rm DI}$
of the derivative interactions in dRGT massive gravity gives rise to the non-minimally coupled Gauss-Bonnet term\footnote{Note that this interaction itself produces 
the second order differential equation of motion. 
However, in the context of massive gravity,
the nonlinear derivative interactions unfortunately 
contain Boulware-Deser ghost \cite{Kimura:2013ika, deRham:2013tfa}.} :
\begin{eqnarray}
\mathcal{L}_{\pi GB}=\mpl^2 \frac{a_4}{\Lambda^3}\pi (R_{\alpha\beta\gamma\delta}R^{\alpha\beta\gamma\delta}-4R_{\alpha\beta}R^{\alpha\beta}+R^2). 
\end{eqnarray}
As it is known,  
Gauss-Bonnet terms can give rise to accelerated expansion so 
that we will now modify the original proxy theory to include this new coupling of the scalar field 
to the Gauss-Bonnet term. Since we construct this additional Gauss-Bonnet term by covariantizing the decoupling limit of the derivative interactions of the dRGT theory, the resulting theory can still be considered as a proxy theory to massive gravity. The additional contributions in the energy density, pressure, and scalar field equation coming from $\mathcal{L}_{\pi GB}$ are
are given by
\begin{eqnarray}
  \rho_{\pi \rm GB}&=& \mpl^2\frac{24a_4}{\Lambda^6}H^3\dot\pi, \\
  p_{\pi \rm GB}&=& -\mpl^2\frac{8a_4}{\Lambda^3}(2H^3{\dot \pi}+2H {\dot H} {\dot \pi}+H^2{\ddot \pi}), \\
  \dot\phi_{\pi \rm GB}&=&-\mpl\frac{8a_4}{\Lambda^3} H.
\end{eqnarray}
The cosmological equations in this case can be expressed as the following autonomous system:
\begin{eqnarray}
\frac{d\hat{y}}{dN}&=&-\frac
{\left(5-13\hat{y}+(9-41c_3)\hat{y}^2+57c_3\hat{y}^3+90c_3^2\hat{y}^4\right)+4\epsilon \hat{H}^2(3-3\hat{y}-10c_3\hat{y}^2)}
{1-6\hat{y}+6(1-5c_3)\hat{y}^2+52c_3\hat{y}^3+105c_3^2\hat{y}^4+16\hat{H}^4+8\epsilon \hat{H}^2(1-2\hat{y}-9c_3\hat{y}^2)}\hat{y}\nonumber\\\nonumber\\
\frac{d\hat{H}}{dN}&=&-\frac
{2-8\hat{y}+(9-33c_3)\hat{y}^2+72c_3\hat{y}^3+135c_3^2\hat{y}^4+16\hat{H}^4+12\epsilon \hat{H}^2(1-2\hat{y}+8c_3\hat{y}^2)}
{1-6\hat{y}+6(1-5c_3)\hat{y}^2+52c_3\hat{y}^3+105c_3^2\hat{y}^4+16\hat{H}^4+8\epsilon \hat{H}^2(1-2\hat{y}-9c_3\hat{y}^2)}\hat{H},\nonumber\\
\label{autonomousGB}
\end{eqnarray}
where $\epsilon\equiv {\rm sign}(b_4)$ and $\hat{H}\equiv H \sqrt{|b_4|}$, with $b_4\equiv a_4M_p^3/\Lambda^3$ (and the number of e-folds is defined with such rescaled Hubble expansion rate). In order to look for critical points with $H\neq0$, we solve for $\hat{H}^2$ from the equation $d\hat{y}/dN=0$ and plug the obtained solution into the equation $d\hat{H}/dN=0$. After doing so, we arrive at the following equation:
\begin{equation}
\frac{\hat{H}}{2-3\hat{y}-10c_3\hat{y}^2}=0,
\end{equation}
whose solution is again $\hat{H}=0$, signaling that the simple coupling of the scalar field to the Gauss-Bonnet term that we have considered is not able to introduce additional critical points. One can clearly see in Fig. \ref{phasemapGB} that there are no additional critical points and that the Minkowski solution is the only attractor solution even if we include the additional Gauss-Bonnet term.

\subsection{Shift symmetry breaking term $\pi R$}
\begin{figure}[h!]
 \begin{center}
 \includegraphics[width=10.cm]{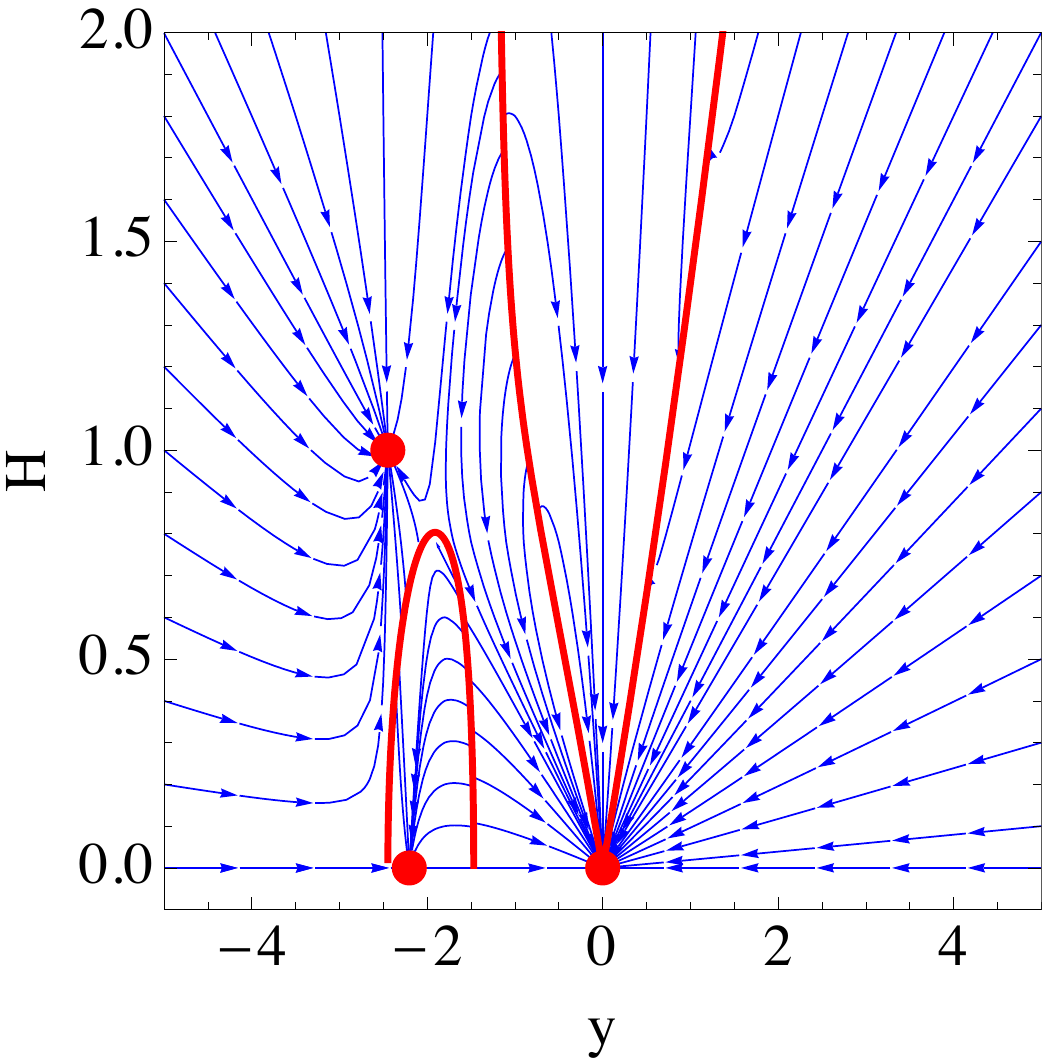}
 \caption{ In this figure we show an example of the phase map for the proxy theory without the $\pi R$ term that spoils the existence of de Sitter critical points. We can see that the de Sitter solution is an attractor of the cosmological evolution. The red lines denote the corresponding separatrices.}
 \label{phasemapnopiR}
 \end{center}
 \end{figure}

Above we have shown that even if we include the additional Gauss-Bonnet term into the proxy theory, which also has its origin from the decoupling limit of massive gravity, the only critical point existing 
in the phase map of the proxy theory is the vacuum Minkowski solution. However, this is not surprising. The problematic term avoiding the existence of de Sitter critical points in the cosmological evolution is the $\pi R$ term in the action. The original approximation $\pi H\ll \dot{\pi}$ used in \cite{deRham:2011by} actually means that exactly this term is negligible. However, our findings show that such a term cannot be consistently maintained small and it is the responsible term for the absence of de Sitter solutions in the proxy theory. Thus, a natural modification of it that will lead to de Sitter solutions consists in simply dropping the problematic term $\pi R$ from the action. It is evident that this modified theory will have de Sitter solutions because in that case the approximation used in \cite{deRham:2011by} is exact. In fact, such a term is the only one violating the shift symmetry so that without it, only the derivatives of the scalar field are physically relevant, but not the value of the field itself. We can proceed analogously as before to obtain the corresponding autonomous system and look for the critical points. When doing so, one can show that there are de Sitter critical points and that they are stable, since the eigenvalues of the matrix determining the linearized system around the de Sitter critical point are both $-3$, confirming the results that have been obtained under the approximation $\pi H\ll \dot{\pi}$ in  \cite{deRham:2011by} . In Fig. \ref{phasemapnopiR} we plot an example of the phase map for the case without the $\pi R$ term in the action and one can indeed see the existence of the de Sitter attractor. The theory without the $\pi R$ term can be considered by its own and represents an interesting subclass of Horndeski interactions. However, its original motivation from massive gravity would be lost. In the context of massive gravity putting $\pi R$ to zero would correspond to putting the kinetic term for the helicity-0 degree of freedom to zero $h^{\mu\nu}X^{(1)}_{\mu\nu}=0$. Thus, this would yield strong coupling issues in the original theory. Since we are only interested in the proxy theory related to massive gravity, we do not consider this option any further. 

\section{Discussion and Summary}

In this paper, we studied the cosmological dynamics of the proxy theory. 
For homogeneous and isotropic universe, there is de Sitter solution found in \cite{deRham:2011by}; 
however we show that this solution can be realized during only transient regime 
and can not be an attractor.
In order to realize this transient de Sitter regime,
we need fine-tuning of the initial conditions of the scalar field,
thus the homogeneous and isotropic universe 
in the proxy theory can not be an alternative theory for dark energy model.
Instead, the space-time approaches Minkowski space-time
or type II singularity at the end, depending on initial conditions.

In the proxy theory, the constant shift symmetry,
$\pi \to \pi +c$, 
is broken by $\pi R$ interaction term while
the decoupling limit theory in massive gravity 
satisfies this symmetry.
If the theory satisfies the shift symmetry,
then the field equation of the scalar field
obeys $\ddot\phi+3H\dot\phi=0$, 
where $\dot \phi$ depends on models.
In this case this equation can be easily solved,
which gives ${\dot \phi} \propto a^{-3}$.
This means that whatever the $\dot\phi$ is
this variable will be diluted in the future,
signaling an attractor solution. 
Furthermore, thanks to the shift symmetry,
$\dot \phi$ only depends on $\dot \pi$, 
and $\pi$ never comes in any equations of motion, which means 
${\dot \pi}={\rm const}$ could be an attractor solution
with the wide range of initial conditions.
This can be applied to the most general second order
scalar-tensor theory which satisfies the shift symmetry.
However, it should be noted that this is just 
a sufficient condition to have de Sitter attractor,
not a neccessary condition. 
The shift symmetry breaking example in the Galileon theory can be found in 
\cite{Silva:2009aa,Kobayashi:2010ab}, 
there exist (quasi-) de Sitter attractor solution in these models.
In addition, the case of massive scalar fields is an exception.

One should note that there is exact de Sitter solution 
in the decoupling limit theory of massive gravity. 
Since the proxy theory shares the same decoupling limit
with massive gravity, 
there should be exact de Sitter attractor solution 
within the patch enclosed by a sphere of radius,
whose domain is order of the current horizon scale $H_0^{-1}$.
This approximate solution should be connected to
inhomogeneous or anisotropic solutions in the proxy theory in a similar way as it is the case in massive gravity itself. However, this would rely on the successful implementation of the Vainshtein mechanism \cite{PhysRevD.84.124046}. It would be interesting to study this kind of inhomogeneous and/or anisotropic solutions in a future work.



\acknowledgments 
L.H. is supported by the Swiss National Science Foundation. 
R.K. is supported in part by a Grant-in-Aid for JSPS Fellows. The research by K.Y. is supported in part by Grant-in-Aid for 
Scientific researcher of Japanese Ministry of Education, 
Culture, Sports, Science and Technology (No.~21540270 
and No.~21244033).


\bibliographystyle{JHEPmodplain}
\bibliography{references}

\providecommand{\href}[2]{#2}\begingroup\raggedright\begin{thebibliography}{10}

\bibitem{FP:1939aa}
W.~P. M.~Fierz, {\it {On Relativistic Wave Equations for Particles of Arbitrary
  Spin in an Electromagnetic Field}},  {\sl Proc. R. Soc.} {\bf A173} (1939)
  211.

\bibitem{Boulware:1972aa}
D.~G. Boulware and S.~Deser, {\it {Can gravitation have a finite range?}},
  {\sl Phys. Rev.} {\bf D6} (1972), no.~3368.

\bibitem{deRham:2010ik}
C.~de~Rham and G.~Gabadadze, {\it {Generalization of the Fierz-Pauli Action}},
  {\sl Phys.Rev.} {\bf D82} (2010) 044020,
  [\href{http://arxiv.org/abs/1007.0443}{{\sf arXiv:1007.0443}}],
  [\href{http://dx.doi.org/10.1103/PhysRevD.82.044020}{{\sf
  doi:10.1103/PhysRevD.82.044020}}].

\bibitem{Rham:2011aa}
C.~de~Rham, G.~Gabadadze, and A.~J. Tolley, {\it Resummation of massive
  gravity},  {\sl Phys.Rev.Lett.} {\bf 106} (2011) 231101,
  [\href{http://arxiv.org/abs/1011.1232}{{\sf 1011.1232}}].

\bibitem{deRham:2012ew}
C.~de~Rham, G.~Gabadadze, L.~Heisenberg, and D.~Pirtskhalava, {\it
  {Nonrenormalization and naturalness in a class of scalar-tensor theories}},
  {\sl Phys.Rev.} {\bf D87} (2013), no.~8 085017,
  [\href{http://arxiv.org/abs/1212.4128}{{\sf arXiv:1212.4128}}],
  [\href{http://dx.doi.org/10.1103/PhysRevD.87.085017}{{\sf
  doi:10.1103/PhysRevD.87.085017}}].

\bibitem{deRham:2013qqa}
C.~de~Rham, L.~Heisenberg, and R.~H. Ribeiro, {\it {Quantum Corrections in
  Massive Gravity}},  {\sl Phys.Rev.} {\bf D88} (2013) 084058,
  [\href{http://arxiv.org/abs/1307.7169}{{\sf arXiv:1307.7169}}],
  [\href{http://dx.doi.org/10.1103/PhysRevD.88.084058}{{\sf
  doi:10.1103/PhysRevD.88.084058}}].

\bibitem{PhysRevD.84.124046}
G.~D'Amico, C.~de~Rham, S.~Dubovsky, G.~Gabadadze, D.~Pirtskhalava, and A.~J.
  Tolley, {\it Massive cosmologies},  {\sl Phys. Rev. D} {\bf 84} (Dec, 2011)
  124046, [\href{http://dx.doi.org/10.1103/PhysRevD.84.124046}{{\sf
  doi:10.1103/PhysRevD.84.124046}}].

\bibitem{Gumrukcuoglu:2011aa}
A.~Gumrukcuoglu, C.~Lin, and S.~Mukohyama, {\it Open frw universes and
  self-acceleration from nonlinear massive gravity},  {\sl JCAP11(2011)030}
  (09, 2011) [\href{http://arxiv.org/abs/1109.3845}{{\sf 1109.3845}}].

\bibitem{Gumrukcuoglu:2012aa}
A.~E. Gumrukcuoglu, C.~Lin, and S.~Mukohyama, {\it Cosmological perturbations
  of self-accelerating universe in nonlinear massive gravity},  {\sl JCAP} {\bf
  03} (2012) 006, [\href{http://arxiv.org/abs/1111.4107}{{\sf 1111.4107}}].

\bibitem{Gumrukcuoglu:2012ab}
A.~E. Gumrukcuoglu, C.~Lin, and S.~Mukohyama, {\it Anisotropic
  friedmann-robertson-walker universe from nonlinear massive gravity},  {\sl
  Phys. Lett.} {\bf B717} (2012) 295,
  [\href{http://arxiv.org/abs/1206.2723}{{\sf 1206.2723}}].

\bibitem{Felice:2012aa}
A.~D. Felice, A.~E. Gumrukcuoglu, and S.~Mukohyama, {\it Massive gravity:
  nonlinear instability of the homogeneous and isotropic universe},  {\sl Phys.
  Rev. Lett.} {\bf 109} (2012) 171101,
  [\href{http://arxiv.org/abs/1206.2080}{{\sf 1206.2080}}].

\bibitem{Rham:2011ab}
C.~de~Rham, G.~Gabadadze, L.~Heisenberg, and D.~Pirtskhalava, {\it Cosmic
  acceleration and the helicity-0 graviton},  {\sl Phys.Rev.D} {\bf 83} (2011)
  103516, [\href{http://arxiv.org/abs/1010.1780}{{\sf 1010.1780}}].

\bibitem{Koyama:2011wx}
K.~Koyama, G.~Niz, and G.~Tasinato, {\it {The Self-Accelerating Universe with
  Vectors in Massive Gravity}},  {\sl JHEP} {\bf 1112} (2011) 065,
  [\href{http://arxiv.org/abs/1110.2618}{{\sf arXiv:1110.2618}}],
  [\href{http://dx.doi.org/10.1007/JHEP12(2011)065}{{\sf
  doi:10.1007/JHEP12(2011)065}}].

\bibitem{Gabadadze:2013aa}
G.~Gabadadze, K.~Hinterbichler, D.~Pirtskhalava, and Y.~Shang, {\it On the
  potential for general relativity and its geometry},  {\sl Phys. Rev. D} {\bf
  88} (2013) 084003, [\href{http://arxiv.org/abs/1307.2245}{{\sf 1307.2245}}].

\bibitem{Leon:2013qh}
G.~Leon, J.~Saavedra, and E.~N. Saridakis, {\it {Cosmological behavior in
  extended nonlinear massive gravity}},  {\sl Class.Quant.Grav.} {\bf 30}
  (2013) 135001, [\href{http://arxiv.org/abs/1301.7419}{{\sf
  arXiv:1301.7419}}],
  [\href{http://dx.doi.org/10.1088/0264-9381/30/13/135001}{{\sf
  doi:10.1088/0264-9381/30/13/135001}}].

\bibitem{deRham:2011by}
C.~de~Rham and L.~Heisenberg, {\it {Cosmology of the Galileon from Massive
  Gravity}},  {\sl Phys.Rev.} {\bf D84} (2011) 043503,
  [\href{http://arxiv.org/abs/1106.3312}{{\sf arXiv:1106.3312}}],
  [\href{http://dx.doi.org/10.1103/PhysRevD.84.043503}{{\sf
  doi:10.1103/PhysRevD.84.043503}}].

\bibitem{Horndeski:1974aa}
G.~Horndeski, {\it {Second-order scalar-tensor field equations in a
  four-dimensional space}},  {\sl Int. J. Theor. Phys.} {\bf 10} (1974),
  no.~363.

\bibitem{Deffayet:2011aa}
C.~Deffayet, X.~Gao, D.~A. Steer, and G.~Zahariade, {\it From k-essence to
  generalised galileons},  \href{http://arxiv.org/abs/1103.3260}{{\sf
  1103.3260}}.

\bibitem{Nicolis:2009aa}
A.~Nicolis, R.~Rattazzi, and E.~Trincherini, {\it The galileon as a local
  modification of gravity},  {\sl Phys.Rev.D} {\bf 79} (2009) 064036,
  [\href{http://arxiv.org/abs/0811.2197}{{\sf 0811.2197}}].

\bibitem{Tasinato:2014eka}
G.~Tasinato, {\it {Cosmic Acceleration from Abelian Symmetry Breaking}},
  \href{http://arxiv.org/abs/1402.6450}{{\sf arXiv:1402.6450}}.

\bibitem{Heisenberg:2014rta}
L.~Heisenberg, {\it {Generalization of the Proca Action}},
  \href{http://arxiv.org/abs/1402.7026}{{\sf arXiv:1402.7026}}.

\bibitem{Burrage:2011bt}
C.~Burrage, C.~de~Rham, and L.~Heisenberg, {\it {de Sitter Galileon}},  {\sl
  JCAP} {\bf 1105} (2011) 025, [\href{http://arxiv.org/abs/1104.0155}{{\sf
  arXiv:1104.0155}}],
  [\href{http://dx.doi.org/10.1088/1475-7516/2011/05/025}{{\sf
  doi:10.1088/1475-7516/2011/05/025}}].

\bibitem{Jimenez:2013qsa}
J.~B. Jim{\'e}nez, R.~Durrer, L.~Heisenberg, and M.~Thorsrud, {\it {Stability
  of Horndeski vector-tensor interactions}},  {\sl JCAP} {\bf 1310} (2013) 064,
  [\href{http://arxiv.org/abs/1308.1867}{{\sf arXiv:1308.1867}}],
  [\href{http://dx.doi.org/10.1088/1475-7516/2013/10/064}{{\sf
  doi:10.1088/1475-7516/2013/10/064}}].

\bibitem{Arkani-Hamed:2003aa}
N.~Arkani-Hamed, H.~Georgi, and M.~D. Schwartz, {\it Effective field theory for
  massive gravitons and gravity in theory space},  {\sl Ann.Phys.} {\bf 305}
  (2003) 96--118, [\href{http://arxiv.org/abs/hep-th/0210184}{{\sf
  hep-th/0210184}}].

\bibitem{Hassan:2011hr}
S.~Hassan and R.~A. Rosen, {\it {Resolving the Ghost Problem in non-Linear
  Massive Gravity}},  {\sl Phys.Rev.Lett.} {\bf 108} (2012) 041101,
  [\href{http://arxiv.org/abs/1106.3344}{{\sf arXiv:1106.3344}}],
  [\href{http://dx.doi.org/10.1103/PhysRevLett.108.041101}{{\sf
  doi:10.1103/PhysRevLett.108.041101}}].

\bibitem{deRham:2011rn}
C.~de~Rham, G.~Gabadadze, and A.~J. Tolley, {\it {Ghost free Massive Gravity in
  the St\'uckelberg language}},  {\sl Phys.Lett.} {\bf B711} (2012) 190--195,
  [\href{http://arxiv.org/abs/1107.3820}{{\sf arXiv:1107.3820}}],
  [\href{http://dx.doi.org/10.1016/j.physletb.2012.03.081}{{\sf
  doi:10.1016/j.physletb.2012.03.081}}].

\bibitem{Ondo:2013wka}
N.~A. Ondo and A.~J. Tolley, {\it {Complete Decoupling Limit of Ghost-free
  Massive Gravity}},  {\sl JHEP} {\bf 1311} (2013) 059,
  [\href{http://arxiv.org/abs/1307.4769}{{\sf arXiv:1307.4769}}],
  [\href{http://dx.doi.org/10.1007/JHEP11(2013)059}{{\sf
  doi:10.1007/JHEP11(2013)059}}].

\bibitem{Charmousis:2011bf}
C.~Charmousis, E.~J. Copeland, A.~Padilla, and P.~M. Saffin, {\it {General
  second order scalar-tensor theory, self tuning, and the Fab Four}},  {\sl
  Phys.Rev.Lett.} {\bf 108} (2012) 051101,
  [\href{http://arxiv.org/abs/1106.2000}{{\sf arXiv:1106.2000}}],
  [\href{http://dx.doi.org/10.1103/PhysRevLett.108.051101}{{\sf
  doi:10.1103/PhysRevLett.108.051101}}].

\bibitem{Kimura:2011qn}
R.~Kimura and K.~Yamamoto, {\it {Constraints on general second-order
  scalar-tensor models from gravitational Cherenkov radiation}},  {\sl JCAP}
  {\bf 1207} (2012) 050, [\href{http://arxiv.org/abs/1112.4284}{{\sf
  arXiv:1112.4284}}],
  [\href{http://dx.doi.org/10.1088/1475-7516/2012/07/050}{{\sf
  doi:10.1088/1475-7516/2012/07/050}}].

\bibitem{Kobayashi:2010aa}
T.~Kobayashi, M.~Yamaguchi, and J.~Yokoyama, {\it G-inflation: inflation driven
  by the galileon field},  {\sl Phys.Rev.Lett.} {\bf 105} (2010) 231302,
  [\href{http://arxiv.org/abs/1008.0603}{{\sf 1008.0603}}].

\bibitem{Amendola:2012ky}
L.~Amendola, M.~Kunz, M.~Motta, I.~D. Saltas, and I.~Sawicki, {\it {Observables
  and unobservables in dark energy cosmologies}},  {\sl Phys.Rev.} {\bf D87}
  (2013) 023501, [\href{http://arxiv.org/abs/1210.0439}{{\sf
  arXiv:1210.0439}}], [\href{http://dx.doi.org/10.1103/PhysRevD.87.023501}{{\sf
  doi:10.1103/PhysRevD.87.023501}}].

\bibitem{Gomes:2013ema}
A.~Gomes and L.~Amendola, {\it {Scaling cosmological solutions with Horndeski
  Lagrangian}},  \href{http://arxiv.org/abs/1306.3593}{{\sf arXiv:1306.3593}}.

\bibitem{Barrow:2004xh}
J.~D. Barrow, {\it {Sudden future singularities}},  {\sl Class.Quant.Grav.}
  {\bf 21} (2004) L79--L82, [\href{http://arxiv.org/abs/gr-qc/0403084}{{\sf
  arXiv:gr-qc/0403084}}],
  [\href{http://dx.doi.org/10.1088/0264-9381/21/11/L03}{{\sf
  doi:10.1088/0264-9381/21/11/L03}}].

\bibitem{Nojiri:2005sx}
S.~Nojiri, S.~D. Odintsov, and S.~Tsujikawa, {\it {Properties of singularities
  in (phantom) dark energy universe}},  {\sl Phys.Rev.} {\bf D71} (2005)
  063004, [\href{http://arxiv.org/abs/hep-th/0501025}{{\sf
  arXiv:hep-th/0501025}}],
  [\href{http://dx.doi.org/10.1103/PhysRevD.71.063004}{{\sf
  doi:10.1103/PhysRevD.71.063004}}].

\bibitem{deRham:2013tfa}
C.~de~Rham, A.~Matas, and A.~J. Tolley, {\it {New Kinetic Interactions for
  Massive Gravity?}},  \href{http://arxiv.org/abs/1311.6485}{{\sf
  arXiv:1311.6485}}.

\bibitem{Kimura:2013ika}
R.~Kimura and D.~Yamauchi, {\it {Derivative interactions in de
  Rham-Gabadadze-Tolley massive gravity}},  {\sl Phys.Rev.} {\bf D88} (2013)
  084025, [\href{http://arxiv.org/abs/1308.0523}{{\sf arXiv:1308.0523}}],
  [\href{http://dx.doi.org/10.1103/PhysRevD.88.084025}{{\sf
  doi:10.1103/PhysRevD.88.084025}}].

\bibitem{Ohara:2014vua}
Y.~Ohara, S.~Akagi, and S.~Nojiri, {\it {Renormalizable theory of massive spin
  two particle and new bigravity}},  \href{http://arxiv.org/abs/1402.5737}{{\sf
  arXiv:1402.5737}}.

\bibitem{Folkerts:2011ev}
S.~Folkerts, A.~Pritzel, and N.~Wintergerst, {\it {On ghosts in theories of
  self-interacting massive spin-2 particles}},
  \href{http://arxiv.org/abs/1107.3157}{{\sf arXiv:1107.3157}}.

\bibitem{Hinterbichler:2013aa}
K.~Hinterbichler, {\it Ghost-free derivative interactions for a massive
  graviton},  {\sl JHEP} {\bf 10} (2013) 102,
  [\href{http://arxiv.org/abs/1305.7227}{{\sf 1305.7227}}].

\bibitem{Silva:2009aa}
F.~P. Silva and K.~Koyama, {\it Self-accelerating universe in galileon
  cosmology},  {\sl Phys.Rev.D} {\bf 80} (2009) 121301,
  [\href{http://arxiv.org/abs/0909.4538}{{\sf 0909.4538}}].

\bibitem{Kobayashi:2010ab}
T.~Kobayashi, H.~Tashiro, and D.~Suzuki, {\it Evolution of linear cosmological
  perturbations and its observational implications in galileon-type modified
  gravity},  {\sl Phys.Rev.D} {\bf 81} (2010) 063513,
  [\href{http://arxiv.org/abs/0912.4641}{{\sf 0912.4641}}].

\end{thebibliography}\endgroup

\end{document}